\documentclass[%
 reprint,
 amsmath,amssymb,
 aps,
]{revtex4-2}

\usepackage{graphicx}% Include figure files
\usepackage{dcolumn}% Align table columns on decimal point
\usepackage{bm}% bold math
\usepackage{textgreek}
\usepackage{color,soul}
\graphicspath{{figures/}}   % defines the path to the folder with figures

\begin{document}

\preprint{APS/123-QED}

\title{Transverse Ultrafast Laser Inscription in Bulk Silicon}

\author{M.~Chambonneau,$^{1,*}$
M.~Blothe,$^{1}$
Q.~Li,$^{1}$
V.~Yu.~Fedorov,$^{2,3}$
T.~Heuermann,$^{1,4}$
M.~Gebhardt,$^{1,4}$
C.~Gaida,$^{5}$
S.~Tertelmann,$^{6}$
F.~Sotier,$^{6}$
J.~Limpert,$^{1,4,7}$
S.~Tzortzakis,$^{2,8,9}$
S.~Nolte$^{1,7}$}

\affiliation{$^{1}$Institute of Applied Physics, Abbe Center of Photonics, Friedrich-Schiller-University Jena, Albert-Einstein-Straße 15, 07745 Jena, Germany.\\
$^{2}$Science Program, Texas A$\&$M University at Qatar, P.O. Box 23874, Doha, Qatar.\\
$^{3}$P. N. Lebedev Physical Institute of the Russian Academy of Sciences, 53 Leninskiy Prospekt, 119991 Moscow, Russia.\\
$^{4}$Helmholtz-Institute Jena, Fröbelstieg 3, 07743 Jena, Germany.\\
$^{5}$Active Fiber Systems GmbH, Ernst-Ruska-Ring 17, 07745 Jena, Germany.\\
$^{6}$Innolas Photonics GmbH, Justus-von-Liebig-Ring 8, 82152 Krailling, Germany.\\
$^{7}$Fraunhofer Institute for Applied Optics and Precision Engineering IOF, Center of Excellence in Photonics, Albert-Einstein-Straße 7, 07745 Jena, Germany.\\
$^{8}$Institute of Electronic Structure and Laser (IESL), Foundation for Research and Technology—Hellas (FORTH), P.O. Box 1527, GR-71110 Heraklion, Greece.\\
$^{9}$Materials Science and Technology Department, University of Crete, 71003 Heraklion, Greece.\\
*maxime.chambonneau@hotmail.fr}

\date{\today}

\begin{abstract}

In-volume ultrafast laser direct writing of silicon is generally limited by strong nonlinear propagation effects preventing the initiation of modifications. By employing a triple-optimization procedure in the spectral, temporal and spatial domains, we demonstrate that modifications can be repeatably produced inside silicon. Our approach relies on irradiation at $\approx 2$-\textmu m wavelength with temporally-distorted femtosecond pulses. These pulses are focused in a way that spherical aberrations of different origins counterbalance, as predicted by point spread function analyses and in good agreement with nonlinear propagation simulations. We also establish the laws governing modification growth on a pulse-to-pulse basis, which allows us to demonstrate transverse inscription inside silicon with various line morphologies depending on the irradiation conditions. We finally show that the production of single-pulse repeatable modifications is a necessary condition for reliable transverse inscription inside silicon.

\end{abstract}

\maketitle

\section{\label{sec:Intro}Introduction}

Ultrafast laser direct writing is a proven technique for in-volume functionalization of dielectrics in a three-dimensional (3D), fast and contactless way \cite{Beresna2014}. Despite its high desirability for applications including photonics, sensors, photovoltaics, ultrafast microelectronics and quantum computing, this technique has no equivalent in monolithic silicon \cite{Chambonneau2021a}. This lack originates from the formation of filaments during the propagation of ultrashort laser pulses in this highly nonlinear material \cite{Kononenko2012,Kononenko2016,Zavedeev2016,Chanal2017,Mareev2020,Chambonneau2020}. As a consequence of these nonlinear propagation effects in silicon, the energy deposition is strongly delocalized to the prefocal region at levels below the breakdown threshold---thus, representing an uttermost challenge for the production of repeatable modifications which are the basic brick for 3D laser direct writing. While different approaches have been devised for writing lines longitudinally (i.e., along the optical axis) in silicon \cite{Chambonneau2016,Pavlov2017,Matthaus2018,Chambonneau2019,Kammer2019,Turnali2019,Alberucci2020}, demonstrations of transverse line inscription (i.e., perpendicular to the optical axis) are seldom, especially in the femtosecond regime. Two strategies were adopted so far for solving this issue. The first one consists of using longer pulses (i.e., in the picosecond or nanosecond regime) to decrease the aforementioned nonlinearities \cite{Ohmura2007,Verburg2014,Tokel2017,Chambonneau2018b,Wang2019,Wang2020b,Richter2020}. The main disadvantage of this approach lies in its thermal nature which limits the degree of control that can be reached on the inscribed structures. In order to reduce the heat-affected zone, a second strategy consists of employing femtosecond pulses with an altered entrance surface, either by depositing an oxide layer \cite{Nejadmalayeri2005}, or by employing oil immersion \cite{Sreenivas2012}. Nevertheless, in addition to the impractical aspect of this strategy, surface alteration is undesirable---if not impossible---in numerous applications.\\

In this article, we demonstrate contactless femtosecond laser transverse inscription inside silicon employing a triple-optimization procedure which combines beneficial effects in the spectral, temporal and spatial domains. The spectral optimization consists of employing femtosecond laser pulses at $\approx 2$-\textmu m wavelength, where the peak in the nonlinear refractive index at this specific wavelength gives rise to enhanced energy deposition \cite{Zavedeev2016,Chambonneau2019a,Richter2020}. The optimization in the time domain relies on the utilization of temporally-distorted 750-fs duration pulses. Temporal imperfections were recently shown to act as a seed inside silicon allowing the material breakdown \cite{Wang2020,Wang2020a}. We propose an additional optimization in the spatial domain which consists of counterbalancing the spherical aberration induced by the focusing lens with the one provoked by the refractive index mismatch at the air-silicon interface, as supported by point spread function calculations. This triple-optimization approach allows us to demonstrate the initiation of repeatable permanent internal modifications with ultrashort laser pulses, in good agreement with nonlinear propagation simulations. The laws governing modification growth after successive irradiations at the same location are then established. This comprehension on the breakdown behavior leads us to devise transverse femtosecond laser inscription of continuous lines in the bulk of silicon. Three distinct line morphologies which depend on the writing speed and the input pulse energy are exhibited. Ultimately, we demonstrate that continuous lines can be solely written in the conditions where the single-pulse modification probability is 100$\%$---thus, shedding light on the necessity for breakdown reignition on a pulse-to-pulse basis during the inscription.

\section{\label{sec:ExpSetup}Experimental arrangement}

\begin{figure}
\includegraphics*[width=\linewidth]{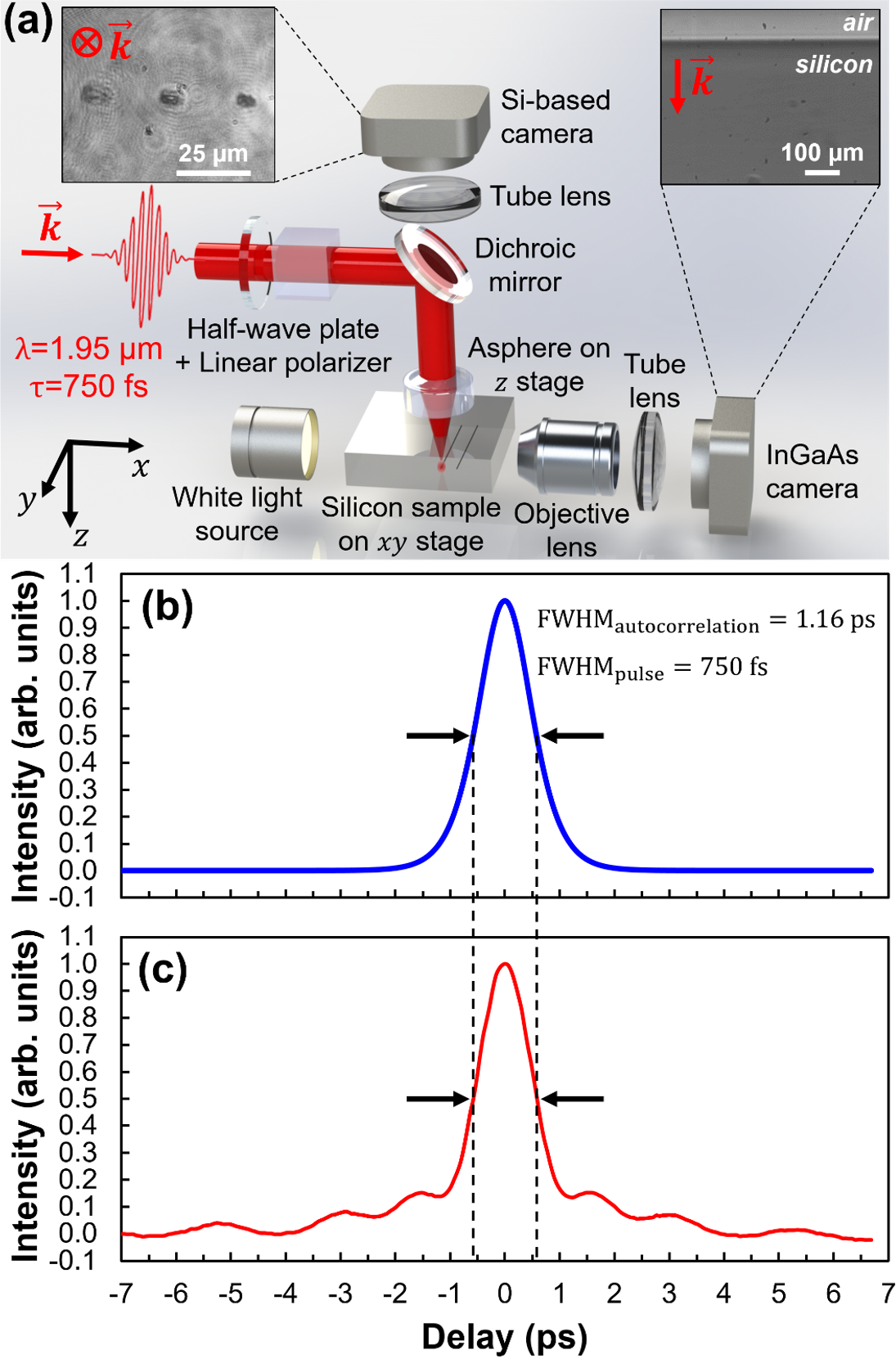}
\caption{\label{fig:FigExperimentalsetup} (a) Schematic of the arrangement employed for internal breakdown and transverse line writing in silicon and associated \textit{in-situ} diagnostics. The vector $\vec{k}$ indicates the direction of laser propagation. The second-order autocorrelation traces in (b) and (c) have been obtained with the two distinct 2-\textmu m laser sources utilized in our study.}
\end{figure}

The experimental arrangement employed for inducing, detecting and characterizing permanent modifications inside silicon is depicted in Fig.~\ref{fig:FigExperimentalsetup}(a). It relies on 750-fs duration pulses [full width at half maximum, assuming a squared hyperbolic secant (sech$^{2}$) temporal shape] at 1.95-\textmu m wavelength emitted at $\Omega_{0}=100$-Hz repetition rate by a Tm-doped fiber laser source. The input pulse energy $E_ {\text{in}}$ is controlled by means of a half-wave plate and a linear polarizer. This energy is measured in air, between the focusing optics and the sample (i.e., the Fresnel reflections are not taken into account), and it may reach $E_ {\text{in}} \approx 2.55$~\textmu J. The beam size before focusing is adjusted by means of a Galilean telescope so that the beam diameter is larger than the diameter of the aspherical lens (Thorlabs, C037TME-D, numerical aperture $\text{NA}=0.85$) employed for focusing the beam inside silicon. This lens is mounted on a stage allowing its displacement along the optical axis~$z$. It is important to note that, while, aspheres usually allow eliminating spherical aberrations, the one we have selected is designed for operation at 9.5-\textmu m wavelength, thus inducing aberrations at our laser wavelength in addition to those induced by the refractive mismatch at the air-silicon interface (this will be discussed below in Sec.~\ref{sec:LinPropag}). The silicon samples (Siegert Wafer, double-side polished, 1-mm thickness, $<100>$-oriented, undoped, 200-\textOmega $\cdot$cm resistivity) can be moved in the $xy$~plane perpendicular to the optical axis~$z$ thanks to a positioning system (Aerotech, ANT 130).\\

Three-dimensional \textit{in-situ} monitoring of the laser-silicon interaction is performed by means of two customized diagnostics. The first one is a transmission microscope relying on white light illumination. The light passes through the sample along the $x$~axis, and images of the modifications in the $yz$~plane are recorded thanks to an objective lens (Mitutoyo, M Plan Apo NIR-HR 50$\times$), a tube lens, and an InGaAs camera (Xenics, Bobcat 320). Images in the $xy$~plane are recorded by a transmission microscope relying on illumination at 1.03-\textmu m (Spectra-Physics, femtoTrain) along the $-z$ direction passing through the silicon sample, the focusing asphere, and a tube lens before being collected by a silicon-based camera. This latter microscope is utilized for determining the relative position of the focus with respect to the sample surface by performing surface damage scans at energies near the breakdown threshold. The focusing depth is obtained by multiplying the displacement $\Delta z$ of the focusing lens along the optical axis starting from the entrance surface by the refractive index of silicon at 1.95-\textmu m ($n_{0} \approx 3.45$ \cite{Li1980}). For improved imaging performance, \textit{post-mortem} characterizations of the modifications are carried out with a transmission infrared phase microscope consisting of a customized Mach-Zehnder interferometer operated at 1.3-\textmu m wavelength whose principle is detailed in Ref.~\cite{Li2016}.\\

Two distinct laser sources have been used in the present study. Both of these sources exhibit similar spectral and spatial features. However, the temporal shapes of the pulses emitted by these laser systems are different. The first laser system (Active Fiber Systems, see details in Ref.~\cite{Baudisch2018}) delivers undistorted bell-shaped pulses as illustrated in Fig.~\ref{fig:FigExperimentalsetup}(b). In stark contrast, as shown by the second-order autocorrelation trace in Fig.~\ref{fig:FigExperimentalsetup}(c), the femtosecond pulses emitted by the second laser source (Innolas Photonics, FEMTO*1950-8-T-2500) exhibit a structured temporal profile. This distortion originates from uncompensated higher-order dispersion resulting in the existence of satellite peaks. However, one has to emphasize that second-order autocorrelation does not allow discriminating if the satellite peaks are pre- or post-pulses.

\section{\label{sec:ResultsDiscussion}Results and discussion}

\subsection{\label{sec:LinPropag} Linear propagation calculations}

Let us first examine the focusing depth allowing to maximize the energy deposition inside silicon, i.e., the spatial optimization. To do so, we have first carried out linear propagation calculations in silicon with our recently developed in-house vectorial model called ``\textit{InFocus}'' \cite{Li2021,InFocus2021}. Briefly, this model gives access to the linear propagation in any medium and relies on point spread function analysis, allowing the determination of the volumetric field distribution of light in the focal region. One specificity of this model is that it accounts not only for the spherical aberration provoked by the refractive index mismatch, but also for the one induced by the focusing lens through its associated Zernike polynomials. While this latter spherical aberration may be neglected for a beam focused with an objective lens, it becomes significant for thick singlet lenses with short focal length. The accuracy of our model to predict these two types of spherical aberrations separately was benchmarked with propagation imaging \cite{Li2021}. While, in principle, similar measurements could be performed in the present study, these would require an objective lens with a numerical aperture higher than the one of the focusing asphere ($\text{NA}=0.85$), which represents a major technical challenge. In the present study, we thus restrict ourselves to linear propagation calculations considering the aspherical lens used in the experiments. As mentioned in Sec.~\ref{sec:ExpSetup}, because of the difference between the design wavelength of this asphere (9.5~\textmu m) and the laser wavelength (1.95~\textmu m), spherical aberration is inevitably induced by the focusing lens.\\

\begin{figure}
\includegraphics*[width=\linewidth]{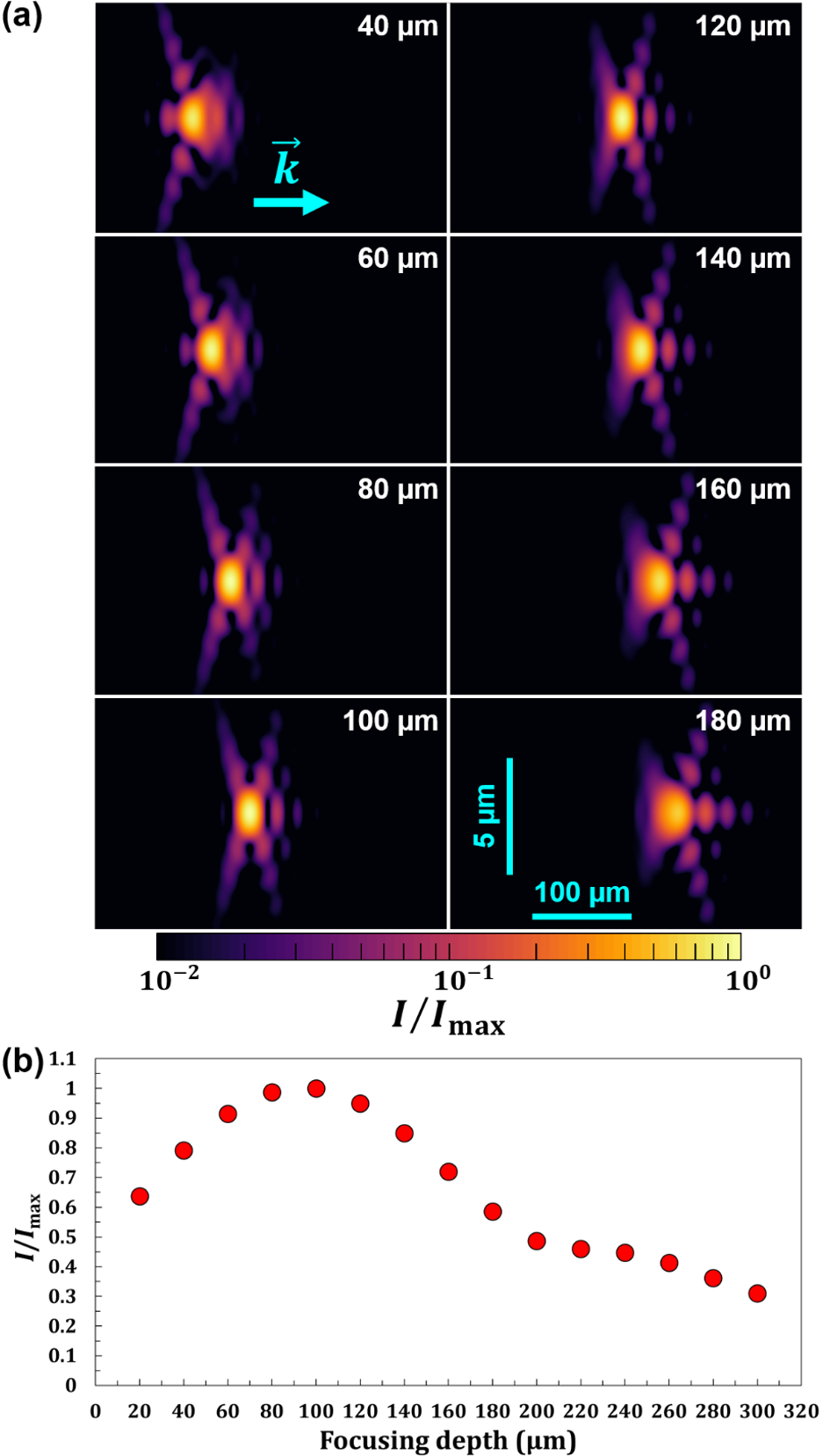}
\caption{\label{fig:LinPropag} (a) Numerical calculations of the on-axis intensity distribution at various focusing depths thanks to our linear propagation model \cite{InFocus2021,Li2021}. The vector $\vec{k}$ indicates the direction of laser propagation. The intensity distributions $I/I_{\text{max}}$ are normalized to the maximum intensity of the whole set of data. The spatial scale applies to all images. (b) Evolution of the normalized maximum intensity for each configuration as a function of the focusing depth.}
\end{figure}

For simplicity, let us first consider that the propagation is linear, i.e., nonlinear effects such as the optical Kerr effect as well as plasma formation and defocusing are neglected. Under this assumption, the results are independent of the temporal profile of the pulses---and, thus, valid for the two employed laser systems [see Figs.~\ref{fig:FigExperimentalsetup}(b) and (c)]. Figure~\ref{fig:LinPropag}(a) shows the intensity distributions for various focusing depths in silicon. For short focusing depths ($\le 80$ \textmu m), the asymmetry in the intensity distributions is ascribable to the predominance of lens-induced spherical aberration. Conversely, for long focusing depths ($\ge 140$ \textmu m), elongated patterns along the propagation direction indicate the predominance of spherical aberration provoked by the refractive index mismatch at the air-silicon interface. Interestingly, for intermediate depths ($\approx 100$ \textmu m), both types of spherical aberrations compensate each other, leading to localized near-symmetric intensity distributions. The maximum intensity reached in each configuration is displayed in Fig.~\ref{fig:LinPropag}(b) and shows a maximum value for 100-\textmu m focusing depth, where both types of spherical aberrations counterbalance. In contrast, for focusing depths $\le 80$ \textmu m and $\ge 120$ \textmu m, the maximum intensity value reached is significantly lower than at $\approx 100$-\textmu m focusing depth. This leads us to the conclusion that the spatial optimization consisting of the compensation of the two types of spherical aberrations is beneficial for provoking breakdown inside silicon. However, one should emphasize that this condition on counterbalancing spherical aberrations taken alone cannot fully explain breakdown experiments as the hypothesis of a linear propagation is unrealistic given our experimental conditions. Indeed, the linear and nonlinear refractive indices of silicon at $\lambda \approx 2$-\textmu m wavelength are $n_{0}=3.45$ \cite{Li1980}, and $n_{2} \approx 9.4 \times 10^{-18}$~m$^{2}\cdot$W$^{-1}$ \cite{Bristow2007,Lin2007,Wang2013}. The corresponding critical power for self-focusing $P_{\text{cr}} \approx 1.8962 \lambda^{2}/\left( 4\pi n_{0} n_{2} \right) \approx 17.7$~kW is much lower than the experimental peak powers ($P \ge 1.7$~MW). Therefore, while spherical aberrations exist experimentally as a linear effect, rigorously, a more sophisticated model accounting for nonlinear propagation is required for determining the optimal focusing depth.

\subsection{\label{sec:NonlinPropag}Nonlinear propagation simulations}

In order to consider the nonlinear propagation inside silicon, we performed numerical simulations based on the unidirectional pulse propagation equation (UPPE) \cite{Kolesik2004,Fedorov2016,Chanal2017}. As the initial condition at the entrance of the silicon sample, we took a pulse with the spatial profile calculated by our vectorial model, which allowed us to include the effect of aberrations into the simulations. The considered temporal shape corresponds to the autocorrelation trace in Fig.~\ref{fig:FigExperimentalsetup}(b). The input pulse energy was chosen to be 1~\textmu J. Together with the UPPE, we solved the following equation for the electron density $\rho$:
\begin{align}
    \frac{\partial\rho}{\partial t} =
        \sigma_2 I^2 (\rho_\text{nt} - \rho) +
        \sigma_\text{B} \frac{\mathcal{E}^2}{U_\text{cr}}
        \frac{\rho_\text{nt} - \rho}{\rho_\text{nt}} \rho -
        \gamma \rho^3 -
        \frac{\rho}{\tau_\text{r}},
\end{align}
where the first term on the right-hand side describes two-photon ionization, second~--- avalanche ionization, third~--- Auger recombination, and fourth~--- electron relaxation. Here $\rho_\text{nt}=5\times10^{28}$~m$^{-3}$ is the density of neutral atoms, $\sigma_2=\beta_2/(2\hbar\omega_0\rho_\text{nt})$ is the cross-section of two-photon ionization with the two-photon absorption coefficient $\beta_2=2.97\times10^{-12}$~m$\cdot$W$^{-1}$~\cite{Bristow2007,Lin2007}, and $I$ is the pulse intensity; $\sigma_\text{B}=(e^2/m_e^*)\nu_c/(\nu_c^2+\omega_0^2)$ is the inverse Bremsstrahlung cross section, where $e$, $m_e$, and $m_e^*=0.5m_e$ are the charge, mass, and effective mass of electron, $\nu_c=0.3\times10^{15}$~s$^{-1}$ is the collisions frequency~\cite{Mouskeftaras2014}, $\omega_0$ is the central pulse frequency, and $\mathcal{E}$ is the pulse electric field, while $U_\text{cr}=1.5U_\text{i}$ with $U_\text{i}=1.12$~eV being the band-gap energy (the coefficient 1.5 accounts for the momentum conservation); $\gamma=1.1\times10^{-42}$~m$^{6}\cdot$s$^{-1}$ is the Auger recombination coefficient~\cite{Jonsson1997}, and finally $\tau_r=100$~fs is the electron relaxation time which we took from Ref.~\cite{Tanimura2019} assuming that, due to collisions, the maximum energy of free electrons cannot be much larger than the band gap. Ultimately, free-carrier diffusion is neglected as this mechanism is not expected to play a major role at the time scale of ultrashort pulses \cite{Mouskeftaras2016}.\\

\begin{figure}
\includegraphics[width=\linewidth]{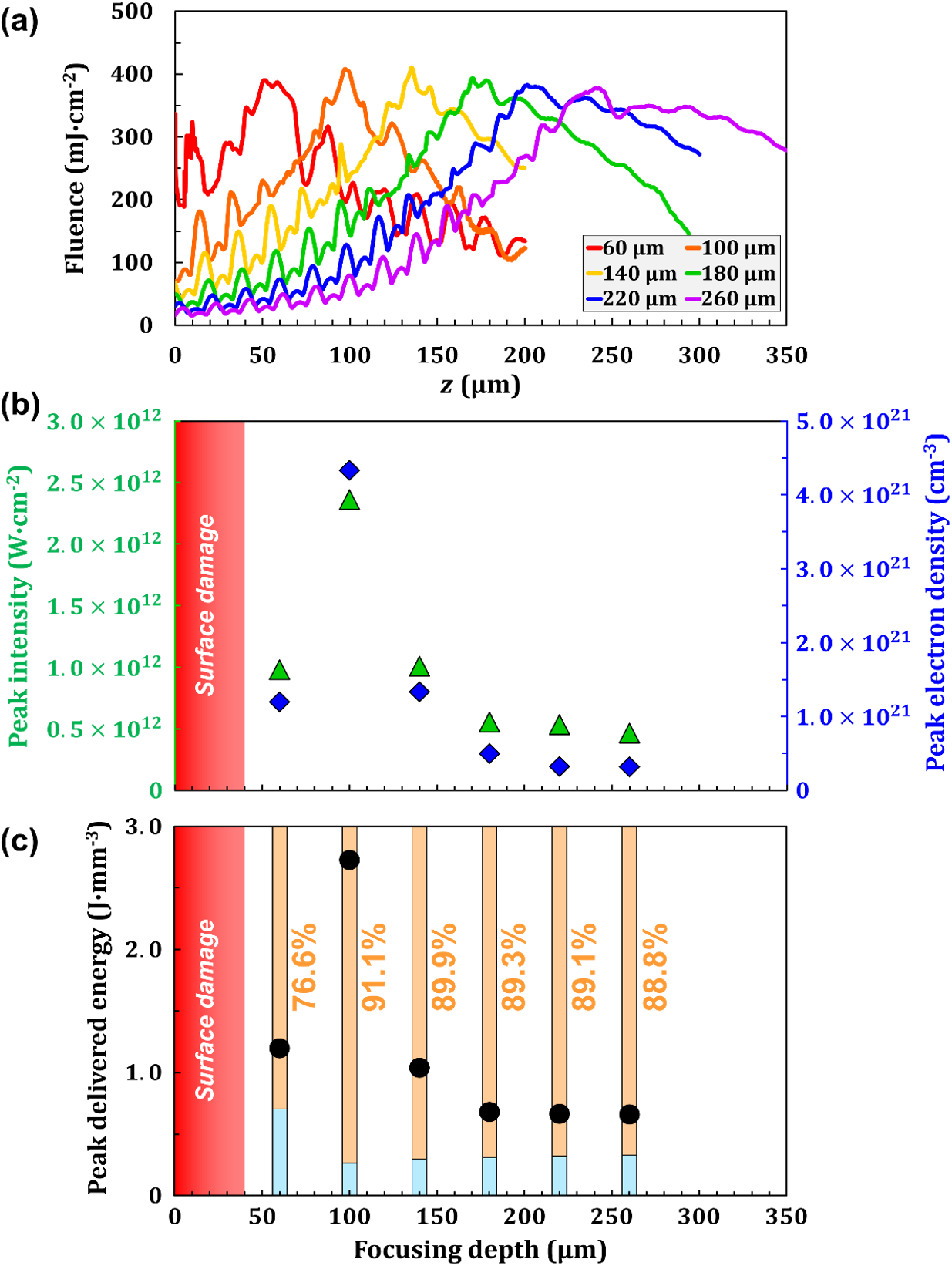}
\caption{\label{fig:FigNonlinearPropagationCalc} Results of numerical simulations of nonlinear propagation in silicon. (a) On-axis fluence profiles at different focusing depths. (b) Evolution of the peak intensity (green) and peak electron density (blue) as a function of the focusing depth. The red region indicates the focusing depths for which surface damage was experimentally detected. (c) Evolution of the peak delivered energy as a function of the focusing depth (black circles). The bars indicate the percentage of the input laser energy spent on ionization (blue) and acceleration (i.e., heating) of electrons (orange, with the indicated percent value).}
\end{figure}

Figure~\ref{fig:FigNonlinearPropagationCalc} summarizes the results of our simulations. Figure~\ref{fig:FigNonlinearPropagationCalc}(a) shows the dependence of the peak fluence on the propagation distance for different focusing depths. Here, one can note that the maximum fluence values are almost independent of the focusing depth. Therefore, based on the fluence data, we cannot conclude that some focal depths are preferable to others. In turn, Fig.~\ref{fig:FigNonlinearPropagationCalc}(b) shows the dependence of peak intensity $I_\text{peak}$ and peak electron density $\rho_\text{peak}$ on the focusing depth. The values of $I_\text{peak}$ and $\rho_\text{peak}$ reach their respective maxima at 100-\textmu m focusing depth, in accordance with the predictions made by our vectorial model (see Fig.~\ref{fig:LinPropag}). However, since the intensity and electron density reach their peak values at some arbitrary point in space (across the beam) and time, one cannot fully rely on this data. Moreover, it is known that criteria based on the peak electron density usually do not provide the most accurate description of breakdown experiments~\cite{Chimier2011}. Therefore, as a more reliable breakdown criterion, we decided to calculate the laser energy delivered to the free-electron subsystem, which then can be transferred to the crystal lattice~\cite{Zhokhov2018} (see the details in Appendix~\ref{sec:AppendixA}). The peak values of the delivered energy (estimated by neglecting the surface damage at depths below 40~\textmu m) are displayed in Fig.~\ref{fig:FigNonlinearPropagationCalc}(c) as a function of the focusing depth. It can be seen that, for 100-\textmu m focusing depth, the peak delivered energy reaches its maximum, indicating that the breakdown probability for this focusing depth is the highest.\\

Additionally, since, in our nonlinear propagation model, the laser energy is only spent on ionization and electron acceleration (heating), by calculating the total laser energy loss and the total energy delivered to free electrons, we can separately find the percent of the initial laser energy spent on ionization and electron heating [see the color bars in Fig.~\ref{fig:FigNonlinearPropagationCalc}(c)]. In particular, we find that, at 100-\textmu m focusing depth, where the peak delivered energy reaches its maximum, the percent of laser energy spent on acceleration of electrons is also the highest and exceeds 90$\%$.\\

\begin{figure}
\includegraphics*[width=\linewidth]{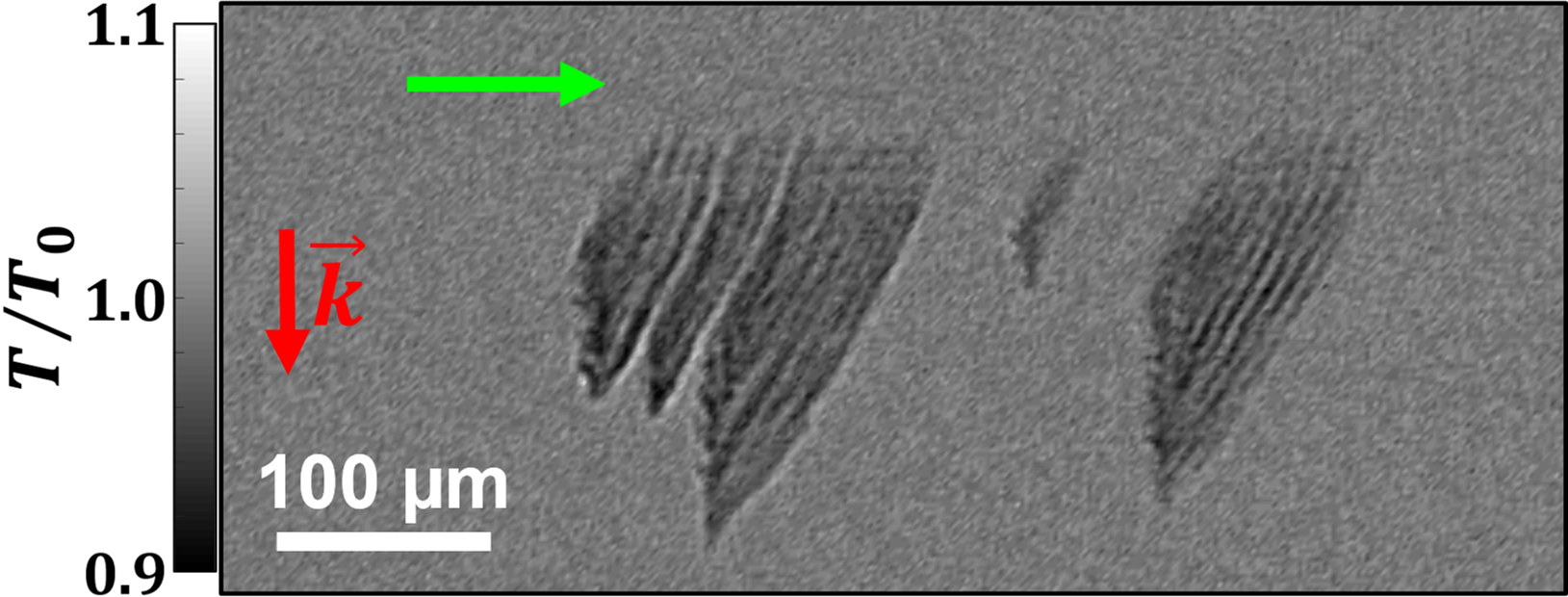}
\caption{\label{fig:FigTemporalShape} Transmission ($T/T_{0}$) image of transverse line inscription attempt with pulses corresponding to the autocorrelation trace in Fig.~\ref{fig:FigExperimentalsetup}(b) at the writing speed $v=20$ \textmu m$\cdot$s$^{-1}$. The vector $\vec{k}$ and the green arrow indicate the direction of laser propagation and the writing direction, respectively.}
\end{figure}

To sum up, both the linear and the nonlinear propagation simulations indicate an optimal focusing depth near 100~\textmu m, which could thus potentially lead to continuous transverse line writing inside silicon. In order to verify experimentally these theoretical predictions, we have first carried out breakdown experiments in the bulk of silicon using the laser source emitting temporally-undistorted pulses [see Fig.~\ref{fig:FigExperimentalsetup}(b)]. These experiments consist of single-pulse irradiation, and the corresponding binary result (i.e., breakdown or not) is evaluated by means of the three-dimensional \textit{in-situ} monitoring system described in Sec.~\ref{sec:ExpSetup}. By repeating the experiments on ten different fresh sites, the bulk modification probability (i.e., the ratio between the number of breakdown sites and the total tested sites) is evaluated. The modification probability lies in the 0--30$\%$ range. In other words, no repeatable modification could be produced in the bulk of silicon. This observation held even by changing the input pulse energy as well as the focusing depth. As illustrated by the endeavored transverse line inscription in Fig.~\ref{fig:FigTemporalShape} where modifications are randomly produced during writing, transverse inscription of continuous lines in this stochastic regime is doomed to failure.  Both the size and the depth of the modifications are unrepeatable. This random behavior for breakdown initiation cannot originate from fluctuations of the laser power due to the high stability ($\approx 1 \%$) of the employed system. It is thus more likely related to the existence of precursor defects inside the material \cite{Bloembergen1974}. One may note the limited spatial expansion of the modification along the writing direction (indicated by the green arrow) in Fig.~\ref{fig:FigTemporalShape}. The modifications can be propagated over $\approx 100$~\textmu m before no more expansion occurs.\\

Transverse inscription of continuous lines could not be achieved despite the benefits expected from irradiations at 2-\textmu m wavelength where the deposited energy is maximized (spectral optimization), and at 100-\textmu m focusing depth where spherical aberrations of two different origins counterbalance (spatial optimization). These negative results lead us to add a third optimization in the temporal domain. In the nanosecond regime, several studies have shown that the temporal pulse shape strongly influences the initiation \cite{Glebov1984,Smith2008,Bataviciute2013,Diaz2014,Lamaignere2017} and the morphology \cite{Chambonneau2014,Chambonneau2018} of laser-produced modifications in fused silica. Recent results showed that the temporal pulse shape is also a key parameter in femtosecond-laser-silicon interaction as perfect Gaussian temporal profile is not suitable for breakdown initiation inside silicon \cite{Wang2020}. Hereafter, the temporal optimization used in the present study consists of employing temporally-distorted pulses associated with the autocorrelation trace in Fig.~\ref{fig:FigExperimentalsetup}(c).

\subsection{\label{sec:SinglePulseProba} Single-pulse breakdown initiation}

\begin{figure}
\includegraphics*[width=\linewidth]{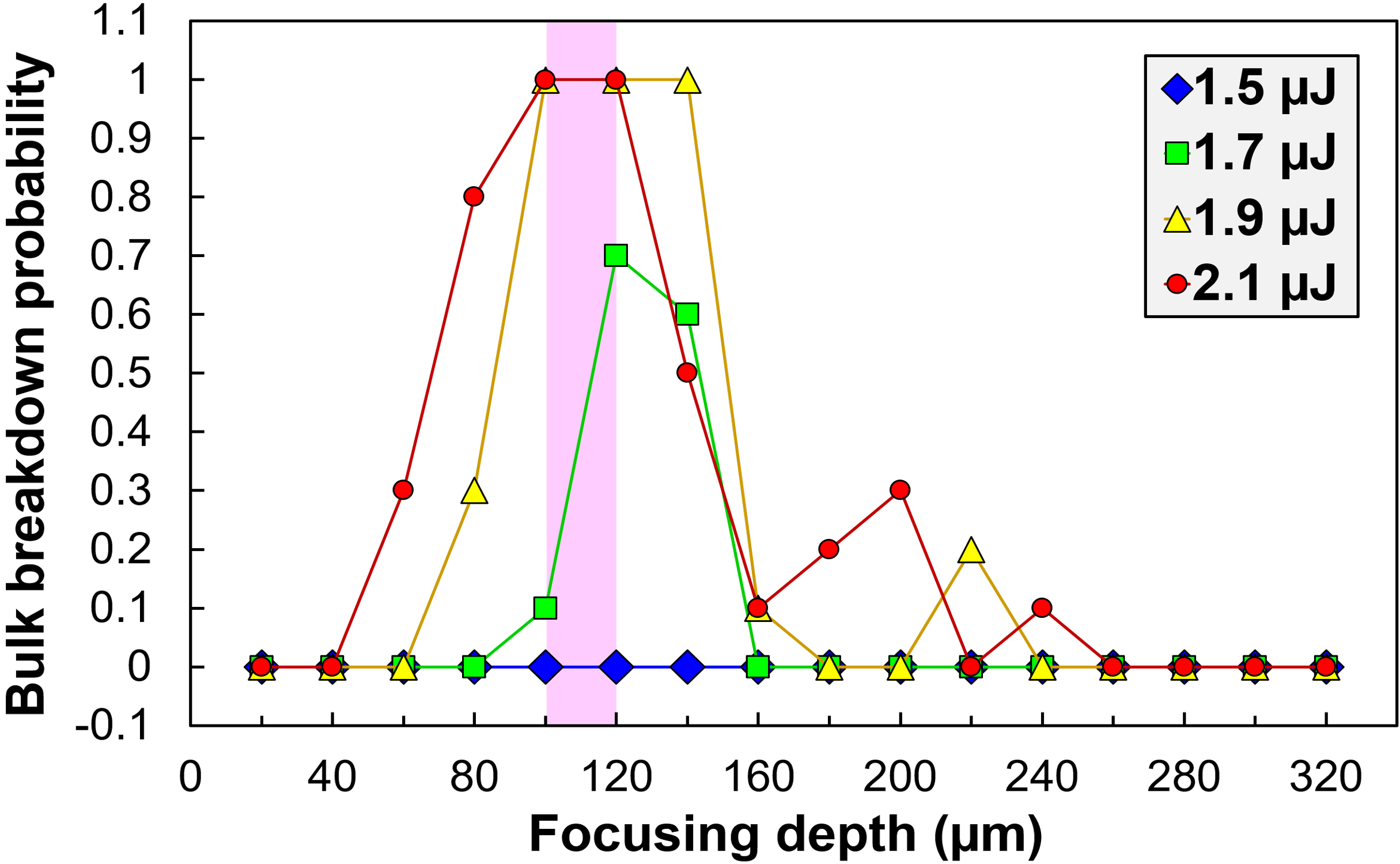}
\caption{\label{fig:FigProbaSinglePulse} Evolution of the bulk breakdown probability inside silicon with single laser pulses as a function of the focusing depth for various input pulse energies. The pink region indicates the depths for which the modification probability reaches 100$\%$ for an input pulse energy $E_{\text{in}} \ge 1.9$~\textmu J.}
\end{figure}

In order to investigate the possibility offered by our triple-optimization procedure for inducing internal modifications in silicon, single-pulse breakdown initiation experiments have been first realized at different focusing depths and input pulse energies. The corresponding results are shown in Fig.~\ref{fig:FigProbaSinglePulse} in terms of bulk breakdown initiation probability as a function of the focusing depth for various input pulse energies $E_{\text{in}}$. Surface damage was systematically detected for focusing depths $\le 40$~\textmu m (not represented in Fig.~\ref{fig:FigProbaSinglePulse}). No bulk breakdown is observed for all focusing depths at $E_{\text{in}}=1.5$~\textmu J, thus defining the modification threshold. The striking feature in Fig.~\ref{fig:FigProbaSinglePulse} is the existence of a narrow window (indicated in pink) for 100--120-\textmu m focusing depth where the breakdown probability shows a maximum value for $E_{\text{in}} \ge 1.7$~\textmu J. This result is in excellent agreement with the theoretical predictions in Secs.~\ref{sec:LinPropag} and \ref{sec:NonlinPropag}. Therefore, this narrow window indicates the focusing depth for which spherical aberrations induced by the lens and the air-silicon interface counterbalance. Interestingly, bulk breakdown probability of 100$\%$ is reached for $E_{\text{in}} \ge 1.9$ \textmu J. This deterministic character suggests that these conditions are well-adapted for ultrafast laser direct writing inside silicon, in opposition to the previous results obtained with temporally-undistorted pulses. Apart from this narrow window, nonzero modification probabilities below 100$\%$ are measured, suggesting non-intrinsic breakdown likely originating from the existence of local precursor defects inside the material \cite{Bloembergen1974}---thus incompatible with the repeatable inscription of functions inside silicon (see Fig.~\ref{fig:FigTemporalShape}).\\

\begin{figure}
\includegraphics*[width=\linewidth]{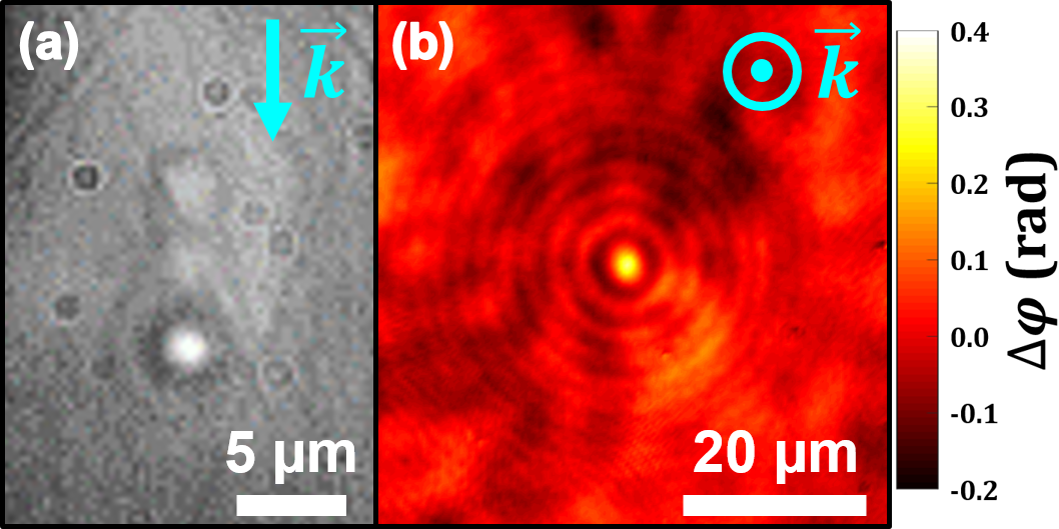}
\caption{\label{fig:MorphoSingleShot} Morphology of a single-pulse modification produced at $E_{\text{in}}=2.1$-\textmu J. (a) \textit{In-situ} transmission microscopy image of the modification in the $yz$~plane. (b) \textit{Post-mortem} phase image of the modification in the $xy$~plane. The phase shift $\Delta \varphi$ is mapped out thanks to a four-step procedure described in Ref.~\cite{Li2016}. The vector $\vec{k}$ indicates the direction of laser propagation.}
\end{figure}

Besides the statistical aspects related to ultrafast laser-induced breakdown inside silicon, we have investigated the associated material changes. To do so, the morphology of single-pulse modifications has been characterized with different methods. First, \textit{in-situ} transverse observations reveal that the modifications consist of a few micro-bubbles along the optical axis~$z$, as shown in Fig.~\ref{fig:MorphoSingleShot}(a). The number of individual bubbles varies from one to four in the tested input energy range, and their typical size is $\approx 4$~\textmu m. \textit{Post-mortem} phase measurements have also been carried out in the $xy$~plane, as shown in Fig.~\ref{fig:MorphoSingleShot}(b). The phase shift---and, thus, the refractive index change---associated with the laser-produced modifications is positive. Two main types of material change may explain such an observation. First, the positive refractive index change may originate from local laser-induced strain inside the material, analogously to breakdown produced in the nanosecond and picosecond regimes \cite{Iwata2018,Chambonneau2018b,Chambonneau2019,Kammer2019}. Second, the modifications may consist of an allotropic change of silicon, from monocrystalline to polycrystalline \cite{Verburg2015}, or other polymorphs \cite{Rapp2015}.

\subsection{\label{sec:MultiPulseDamageGrowth} Multi-pulse modification growth}

\begin{figure}
\includegraphics*[width=0.8\linewidth]{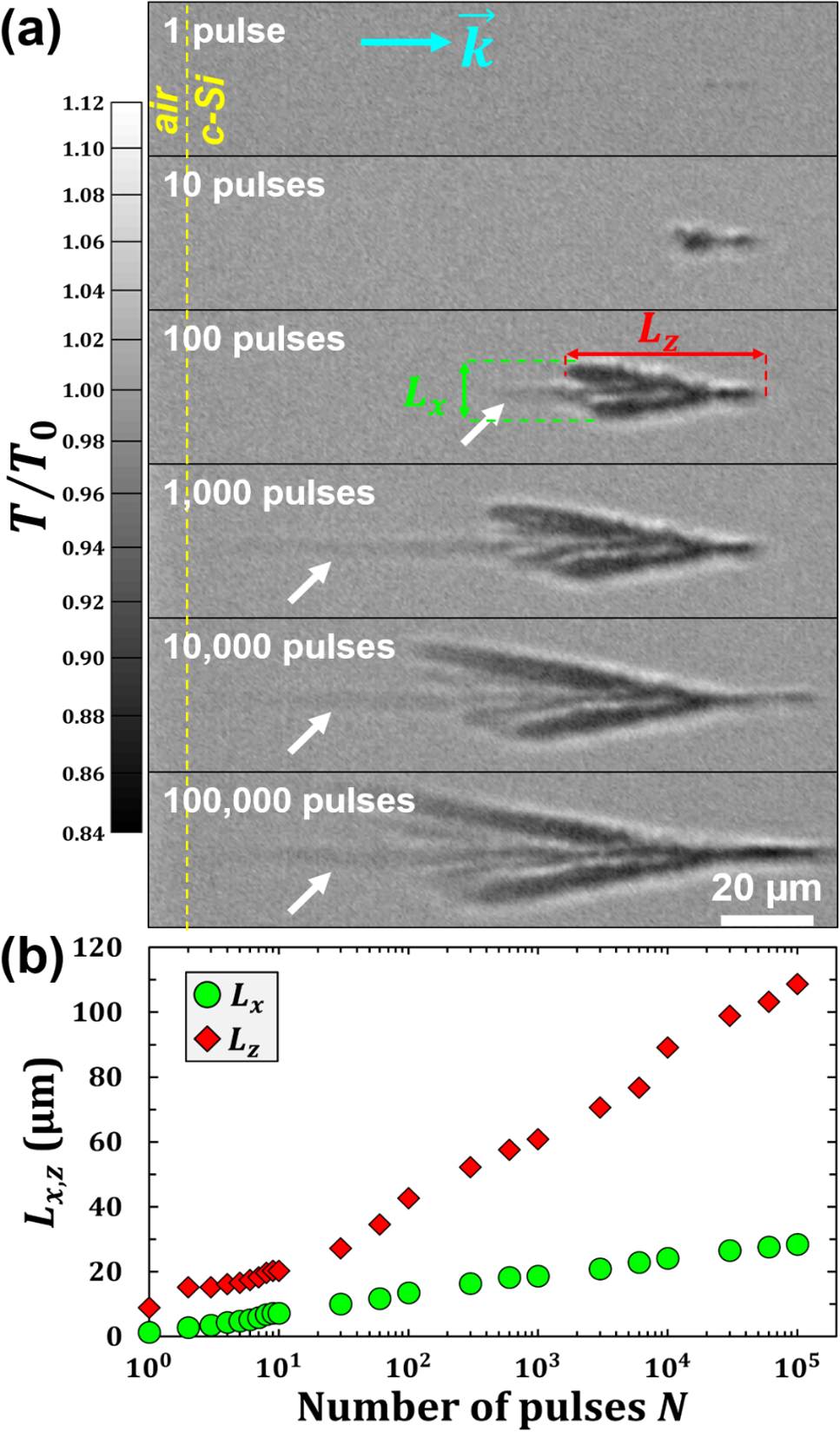}
\caption{\label{fig:Growth} Laser-induced modification growth sequence inside silicon for $E_{\text{in}}=2.1$ \textmu J. (a) Transmission images of a modification for different numbers of irradiations at the same location (120-\textmu m below the entrance surface of the sample). The transmission $T/T_{0}$ is obtained by dividing a transmission optical image after $N$ pulses at the same location by a reference image. The spatial scale applies to all images. The vector $\vec{k}$ indicates the direction of laser propagation. The dotted yellow lines indicate the sample entrance surface. (b) Evolution of the modification length $L_{x}$ and $L_{z}$ indicated in (a) along the $x$ and the $z$~axis, respectively, as a function of the number of applied pulses.}
\end{figure}

Establishing the laws governing modification growth (i.e., the increase in the modification size on a pulse-to-pulse basis) in silicon is a prerequisite for envisioning multi-pulse functionalization processes. While, previously, we have determined the conditions for which spherical aberrations induced by the lens and the air-silicon interface counterbalance, leading to repeatable breakdown (see Fig.~\ref{fig:FigProbaSinglePulse}), experiments on the evolution of the modification size after successive irradiations have been carried out. Such a growth sequence is illustrated by transmission microscopy images in the $xz$~plane after different numbers of applied pulses in Fig.~\ref{fig:Growth}(a). The first pulse focused at 120-\textmu m depth initiates a modification similar to the one in Fig.~\ref{fig:MorphoSingleShot}. This barely detectable single-shot modification expands spatially after several irradiations to form a strongly absorbing damage site. The modification preferentially grows toward the entrance surface of the sample, and only marginally after the focus. This behavior is ascribable to enhanced absorption in the modified zone. One may note the formation of a tail growing rapidly toward the entrance surface for a number of pulses $N \ge 30$, as indicated by white arrows in Fig.~\ref{fig:Growth}(a). While deeper investigations out of the scope of the present study are necessary for establishing the physical origin of this tail, it must be emphasized that the expansion of this tail does not provoke any detectable surface damage. Quantitative analyses of the evolution of the modification lengths $L_{x}$ and $L_{z}$ indicated in Fig.~\ref{fig:Growth}(a) along the $x$ and the $z$~axis, respectively, are shown in Fig.~\ref{fig:Growth}(b) as a function of the number of pulses $N$. For all $N$ values, $L_{z} > L_{x}$, meaning that the modifications are more elongated along the optical axis than in the transverse direction. Both $L_{x}$ and $L_{z}$ scale logarithmically with the number of pulses $N$. Remarkably, this evolution for bulk modifications is slower than for surface modifications in fused silica, which usually exhibit linear and exponential growth on the entrance and the exit surface, respectively \cite{Norton2006,Sozet2016,Chambonneau2018a}.\\

\begin{figure}
\includegraphics*[width=0.9\linewidth]{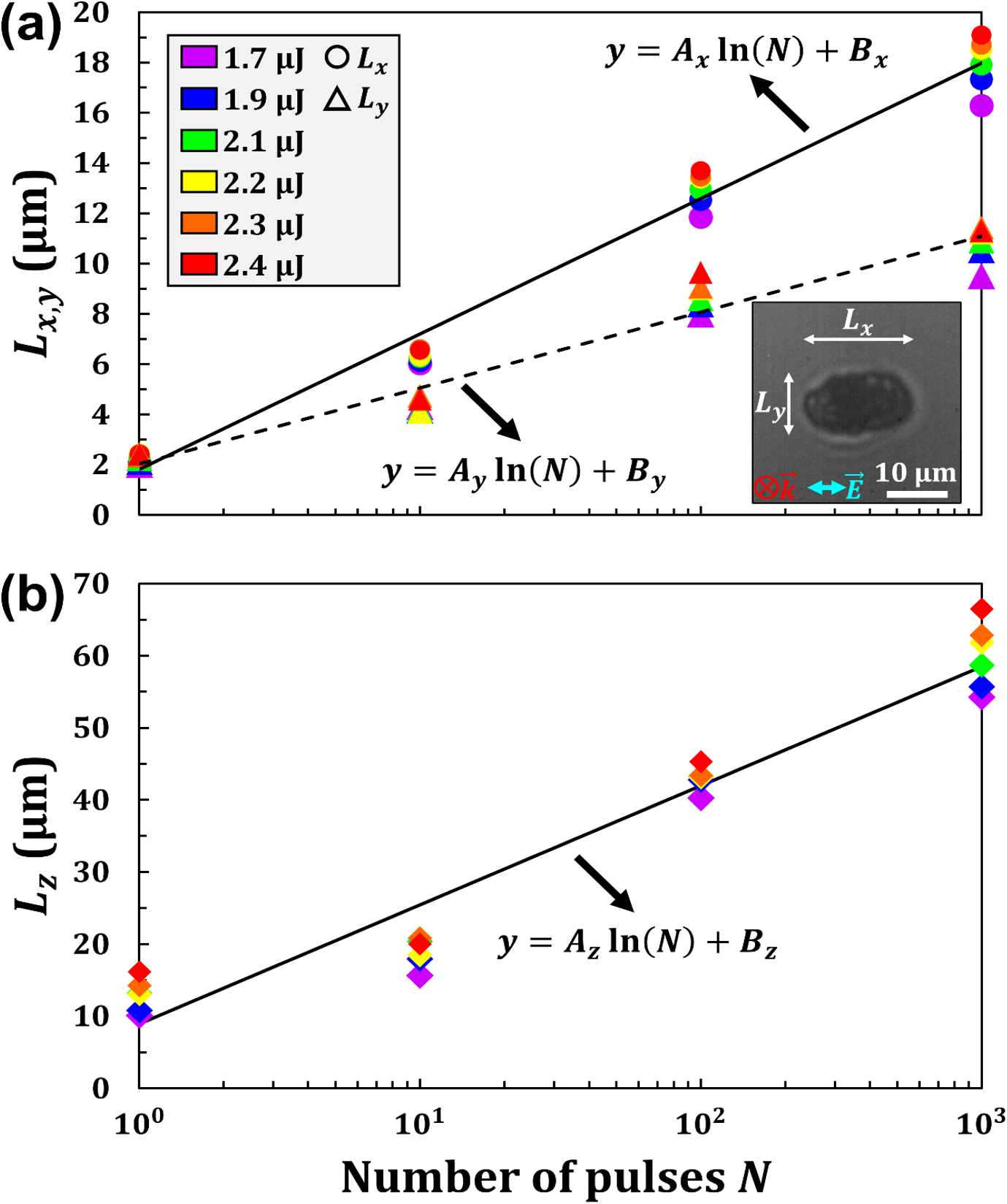}
\caption{\label{fig:GrowthLaws} Laws of modification growth inside silicon. (a) Evolution of the transverse modification dimensions $L_{x}$ (circles) and $L_{y}$ (triangles) along the $x$ and the $y$~axis, respectively, as a function of the number of applied pulses $N$ for different input pulse energies. The inset is an optical microscopy observation of a modification produced after 1,000 irradiations at $E_{\text{in}}=2.4$ \textmu J. The vectors $\vec{k}$ and $\vec{E}$ indicate the direction of laser propagation and the polarisation, respectively. (b) Evolution of the modification dimension $L_{z}$ along the optical axis $z$ for the same input pulse energies as in (a). In both (a) and (b), the curves with indicated equations are logarithmic fits of the experimental data according to Eq.~(\ref{eq:growthfit}). The values of the coefficients $A_{i}$ and $B_{i}$ ($i=x,y$ or $z$) are reported in Table \ref{tab:TableGrowthLaws}.}
\end{figure}

Additional growth experiments have been carried out for different input pulse energies. The corresponding results are displayed in Fig.~\ref{fig:GrowthLaws}. The evolution of the modification lengths $L_{x,y}$ and $L_{z}$ are shown in Figs.~\ref{fig:GrowthLaws}(a) and \ref{fig:GrowthLaws}(b), respectively, as a function of the number of pulses $N$. Similar to the trends in Fig.~\ref{fig:Growth}(b), each data set is fairly well fitted by a logarithmic function as
\begin{equation}
\label{eq:growthfit}
L_{i} = A_{i} \text{ln}(N)+B_{i},
\end{equation}
where $i=x$, $y$ or $z$, and $A_{i}$ and $B_{i}$ are the growth coefficient and the single-pulse modification length along $i$, respectively. Given that the characterizations in Fig.~\ref{fig:MorphoSingleShot} have revealed symmetric modifications in the $xy$~plane, one could intuitively expect a conservation of this symmetry in the multi-pulse configuration---and, thus, $A_{x}=A_{y}$. As shown in Fig.~\ref{fig:GrowthLaws}(a), this is not the case, and the modifications are more elongated along $x$ than along $y$. Consecutive irradiations thus produce elliptical modifications whose major and minor axes are parallel and perpendicular to the laser polarization, respectively, as illustrated by the inset in Fig.~\ref{fig:GrowthLaws}(a) where the modification is 60$\%$ smaller along $y$ than along $x$. Interestingly, elliptical modifications were also observed in silicon for multi-pulse picosecond irradiations in the bulk \cite{Mori2015a,Wang2020,Chambonneau2019a,Das2020}, as well as for nanosecond irradiations on the exit surface \cite{Luong2020}. However, the physical explanation for the origin of this elliptical morphology remains open and requires a dedicated study.\\

The relatively small dispersion in the $L_{x,y,z}$ values in Fig.~\ref{fig:GrowthLaws} for different input pulse energies $E_{\text{in}}$ indicate that the laws of modification growth described by Eq.~(\ref{eq:growthfit}) do not depend much on this parameter in the tested input pulse energy range ($E_{\text{in}}=1.7$--2.4 \textmu J). This is in agreement with the saturation of the laser intensity during the propagation of ultrashort pulses in silicon \cite{Kononenko2016,Zavedeev2016}. A statistical analysis of $A_{i}$ and $B_{i}$ coefficients ($i=x$, $y$ or $z$) appearing in Eq.~(\ref{eq:growthfit}) is recapitulated in Table \ref{tab:TableGrowthLaws}. Consistent with Figs.~\ref{fig:Growth} and \ref{fig:GrowthLaws}, $A_{z}>A_{x,y}$ and $B_{z}>B_{x,y}$, showing again that the modifications grow preferentially along $z$.

\begin{table*}\caption{\label{tab:TableGrowthLaws} Statistical analysis of the $A_{i}$ and $B_{i}$ coefficients ($i=x$, $y$ or $z$) appearing in the laws of modification growth described in Eq.~(\ref{eq:growthfit}). The data have been obtained by analyzing 288 modifications initiated at 120-\textmu m focusing depth by 1, 10, 100 and 1,000 laser pulses with $E_{\text{in}}=1.7$--2.4~\textmu J.}
\begin{ruledtabular}
\begin{tabular}{ccccccc}
&$A_{x}$~(\textmu m)&$B_{x}$~(\textmu m)&$A_{y}$~(\textmu m)&$B_{y}$~(\textmu m)&$A_{z}$~(\textmu m)&$B_{z}$~(\textmu m)\\
\hline
Average & 2.34 & 1.81 & 1.31 & 2.03 & 7.18 & 8.91\\
Standard deviation & 0.12 & 0.07 & 0.08 & 0.13 & 0.31 & 1.43\\
Minimum & 2.12 & 1.71 & 1.14 & 1.86 & 6.82 & 6.49\\
Maximum & 2.48 & 1.90 & 1.38 & 2.25 & 7.66 & 10.54
\end{tabular}
\end{ruledtabular}
\end{table*}

\subsection{\label{sec:TransverseInscription} Transverse inscription}

\begin{figure}
\includegraphics*[width=\linewidth]{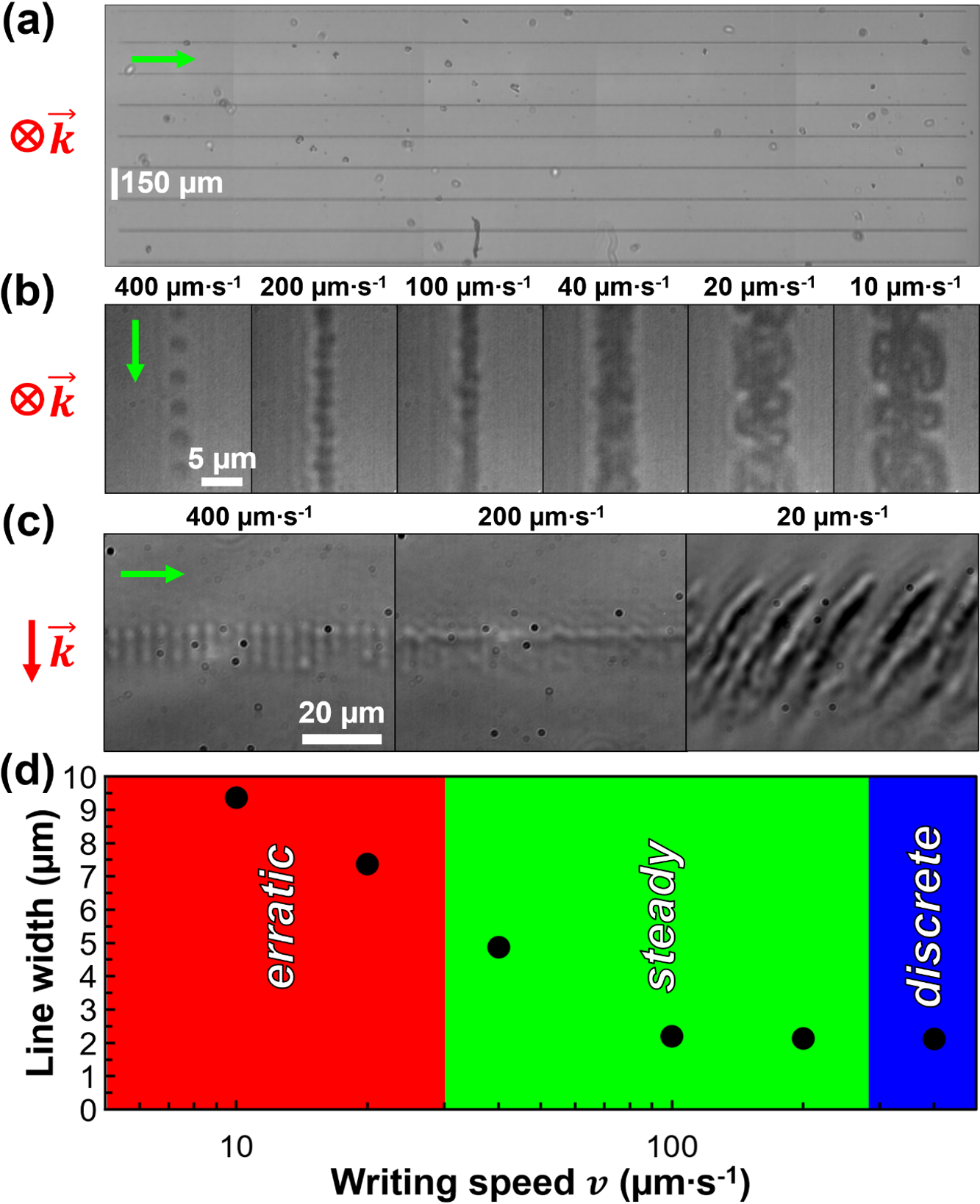}
\caption{\label{fig:FigTransverseLines} Transverse ultrafast-laser-inscribed lines inside silicon. (a) Overview of 4-mm long laser-written lines at 120-\textmu m focusing depth. (b) \textit{Post-mortem} transmission microscopy images in the $yz$~plane of lines inscribed at $E_{\text{in}}=2.5$~\textmu J for different writing speeds $v$. (c) \textit{In-situ} transmission microscopy images in the $xy$~plane of lines inscribed at $E_{\text{in}}=2.5$~\textmu J for different writing speeds $v$. In each sub-figure, the spatial scale applies to all images, the vector $\vec{k}$ indicates the direction of laser propagation, and the green arrow shows the writing direction. (d) Evolution of the line width in the $xy$~plane as a function of $v$ for $E_{\text{in}}=2.5$~\textmu J. The different colors correspond to the indicated morphology domains.}
\end{figure}

By constantly moving the sample perpendicularly to the laser beam at various writing speeds $v$, transverse inscription has been endeavored inside silicon at 120-\textmu m focusing depth and $E_{\text{in}}=2.5$~\textmu J. As shown by the overview in Fig.~\ref{fig:FigTransverseLines}(a), our triple-optimization procedure allows the first demonstration of contactless transverse inscription of lines in bulk silicon with femtosecond laser pulses. Finer characterizations of the line morphology in different imaging planes are displayed in Figs.~\ref{fig:FigTransverseLines}(b) and \ref{fig:FigTransverseLines}(c), revealing three distinct morphologies. The first one appearing for $v=400$~\textmu m$\cdot$s$^{-1}$ is the discrete morphology, consisting of a track of individual single-pulse modifications. This morphology associated with high writing speeds is in excellent agreement with the prediction of the phenomenological model detailed in Appendix \ref{sec:AppendixB}, which allows the determination of the threshold minimum speed value for discrete line inscription $v_{\text{min}} \approx 262$~\textmu m$\cdot$s$^{-1}$. In stark contrast, for $v \le v_{\text{min}}$ the lines are continuous. In particular, for $40 \le v \le 200$~\textmu m$\cdot$s$^{-1}$, the line morphology is steady, i.e., with a nearly constant track width. Finally, for low speeds ($v \le 20$~\textmu m$\cdot$s$^{-1}$), bulgy clusters are exhibited, and the line morphology is erratic (i.e., the track width is not constant). This complex morphology likely originates from cumulative effects as suggested in Fig.~\ref{fig:FigTransverseLines}(c) for $v=20$~\textmu m$\cdot$s$^{-1}$. In this case, the high number of pulses per point ($N \geq 10$) associated with the slow inscription process gives rise to modification growth, provoking shielding and scattering effects at the focus during the subsequent irradiations. As a consequence, the energy deposition during the subsequent irradiations is drastically reduced so that no new modification is initiated at the focus until the growth-induced shielding and scattering effects stop. Finally, the evolution of the line width in the $xy$~plane is displayed in Fig.~\ref{fig:FigTransverseLines}(d) as a function of the writing speed. The line width is the same for $v \ge 100$~\textmu m$\cdot$s$^{-1}$ (i.e., $N \le 2$), and corresponds to the dimensions of single-pulse modifications (see Fig.~\ref{fig:MorphoSingleShot}). This demonstrates the absence of cumulative effects for high writing speeds. In contrast, for $v<100$~\textmu m$\cdot$s$^{-1}$, the line width increases logarithmically as the writing speed decreases, thus defining the threshold speed for the aforementioned cumulative effects.\\

\begin{figure}
\includegraphics*[width=\linewidth]{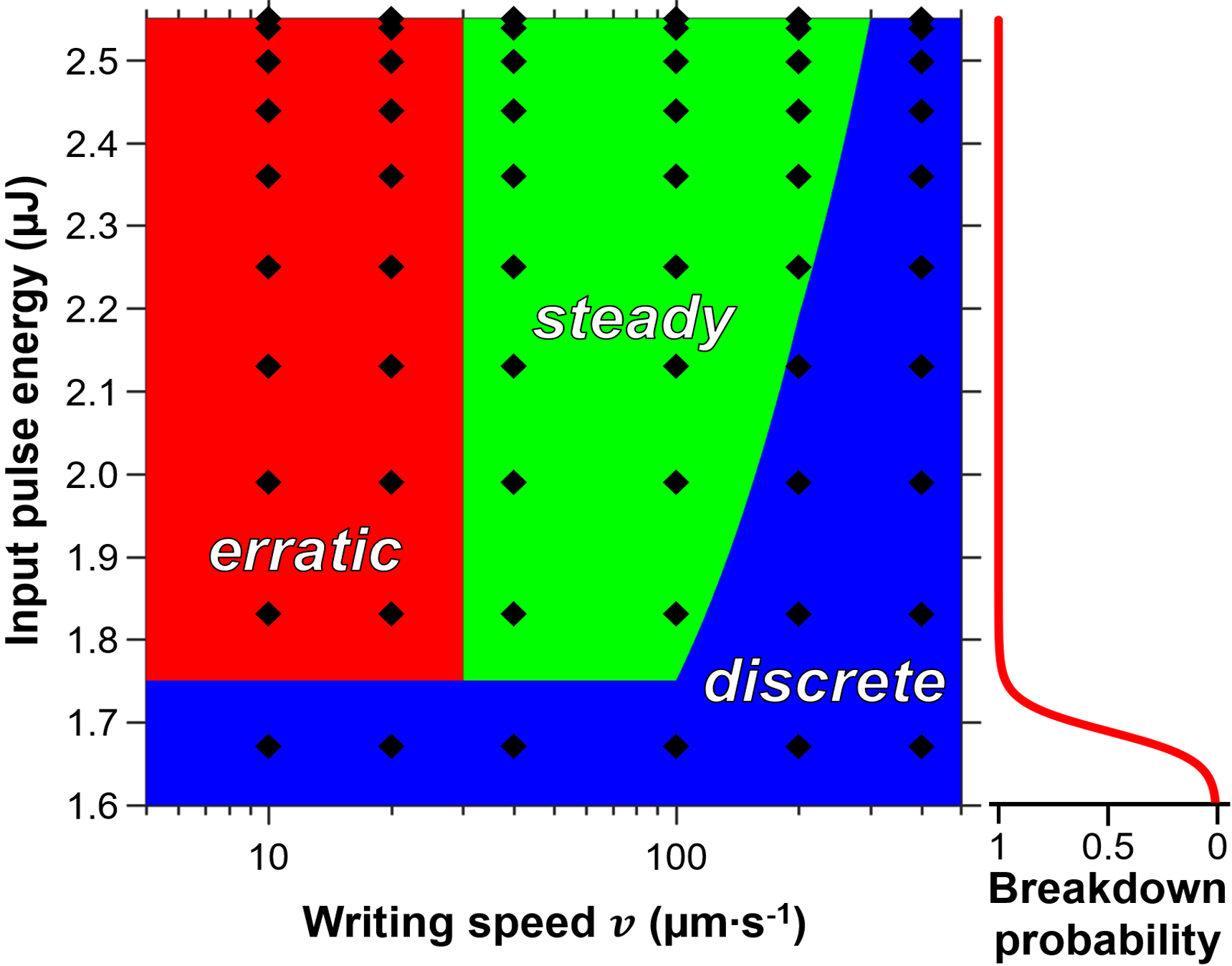}
\caption{\label{fig:FigLineMorphoDomains} Morphology domains of transversely-inscribed lines as a function of the input pulse energy and the writing speed. The focusing depth is 120~\textmu m. The black points indicate the conditions for which line inscription has been endeavored, and the color zones correspond to the different morphology domains. The red curve corresponds to a sigmoid fit of the single-pulse breakdown probability data in Fig.~\ref{fig:FigProbaSinglePulse}.}
\end{figure}

Ultimately, we have studied the energy dependence of the three different morphologies exhibited in Fig.~\ref{fig:FigTransverseLines} (i.e., discrete, steady and erratic) by characterizing transversely-inscribed lines for various input pulse energies and writing speeds. The corresponding results are displayed in Fig.~\ref{fig:FigLineMorphoDomains}. As discussed previously, high input pulse energies lead to discrete and erratic morphology for excessively high and low speeds, respectively. Interestingly, for intermediate speeds, the steady morphology solely exists for $E_{\text{in}}>1.8$~\textmu J, corresponding to a single-pulse breakdown probability of 100$\%$ (see the red curve in Fig.~\ref{fig:FigLineMorphoDomains}). Conversely, the line morphology is discrete for $v=40$--100 \textmu m$\cdot$s$^{-1}$ and $E_{\text{in}} \approx 1.7$ \textmu J, which corresponds to a nonzero modification probability strictly lower than $100 \%$ (see Fig.~\ref{fig:FigProbaSinglePulse}). Given that modification growth occurs at $E_{\text{in}} \approx 1.7$ \textmu J (see Fig.~\ref{fig:GrowthLaws}), one can conclude that continuous line inscription is not related to the reignition of modifications enabling their transverse expansion, but to intrinsic material breakdown. In other words, a single-pulse modification probability of $100 \%$ is a necessary condition for the transverse inscription of continuous lines (either steady or erratic). This conclusion is all the more supported by the observations in Fig.~\ref{fig:FigTemporalShape}.

\section{\label{sec:Summary}Summary}

We have demonstrated the possibility to contactlessly inscribe continuous transverse lines in bulk silicon with ultrafast laser pulses. While this achievement is usually unattainable due to nonlinear propagation effects, we have proposed a triple-optimization procedure for circumventing these limitations. This approach consists of using femtosecond pulses (i) at the beneficial wavelength of $\approx 2$~\textmu m, (ii) exhibiting temporal distortions, and (iii) focused at a depth for which spherical aberrations of different origins counterbalance. The concomitance of these spectral, temporal and spatial optimizations was shown to be indispensable for producing repeatable breakdown---and, thus, transverse inscription---inside silicon.\\

Our methodology first relies on the identification of the conditions that allow obtaining $100 \%$ breakdown probability with single-pulse irradiations. Deterministic breakdown was observed for input pulse energies $E_{\text{in}} \ge 1.9$~\textmu J and focusing depths $\approx 100$--120~\textmu m below the entrance surface. This narrow focusing depth window is in good agreement with point spread function analyses as well as nonlinear propagation simulations. In addition, we have studied the spatial expansion of the modifications on a pulse-to-pulse basis. By applying different numbers of pulses for various input pulse energies, we have determined the laws governing modification growth. In all spatial directions, the growth behavior was found to scale logarithmically with the number of pulses applied, with no strong dependence on the input pulse energy. Finally, lines have been transversely inscribed in the bulk of silicon. Depending on the writing speed and the input pulse energy, the transversely-written lines may show an erratic, steady or discrete morphology. The discrete morphology appears either for input pulse energies leading to single-pulse breakdown probabilities $< 100 \%$, or for too high writing speeds.\\

The demonstration of transversely-inscribed lines paves the way to contactless 3D laser direct writing inside silicon. While we have concentrated our efforts on transverse line inscription at a single focusing depth (120~\textmu m), we anticipate that phase-control elements (e.g., spatial light modulator) can be employed for precompensating spherical aberrations at arbitrary depths---thus, allowing ultrafast laser inscription in all three spatial directions. In the future, a broad range of in-chip applications can be addressed, including optical functionalization, wafer dicing, and microfluidics.

\begin{acknowledgments}
This research has been supported by the Bundesministerium für Bildung und Forschung (BMBF) through the NUCLEUS project, grant No. 03IHS107A, as well as the glass2met project, grant No. 13N15290. It has also been supported by the National Priorities Research Program grant No. NPRP11S-1128-170042 from the Qatar National Research Fund (member of The Qatar Foundation). The authors gratefully acknowledge R. Herda and A. Zach (Toptica Photonics) for technical support.
\end{acknowledgments}

\appendix

\section{\label{sec:AppendixA}Calculations of the delivered energy}

In order to calculate the laser energy delivered to the free-electron subsystem, we first consider the work $W$ done by the electromagnetic field force $\vec{F}$ acting on the electrons of total charge $q$ during time interval $dt$ \cite{Griffiths2013}:
\begin{align}
    W = \vec{F} \cdot \vec{dl}
      = q (\vec{\mathcal{E}} + \vec{v}\times\vec{\mathcal{B}}) \cdot
        (\vec{v} \, dt)
      = q (\vec{\mathcal{E}} \cdot \vec{v}) dt.
\end{align}
where $\vec{\mathcal{E}}$ and $\vec{\mathcal{B}}$ are the electric and magnetic fields, respectively, and $\vec{v}$ is the electron velocity. For continuous charge distribution $q=\rho \, dV$, with electron density $\rho$ and volume element $dV$, we obtain
\begin{align}
    W & = (\rho \, dV) (\vec{\mathcal{E}} \cdot \vec{v}) dt
        = \vec{\mathcal{E}} \cdot (\rho \vec{v}) \, dV dt \notag \\
      & = (\vec{\mathcal{E}} \cdot \vec{J}) \, dV dt,
\end{align}
where $\vec{J}=\rho\vec{v}$ is the current. Thus, the work done by the electromagnetic field force per unit time, per unit volume, that is, the energy delivered to the free electrons per unit time, per unit volume, is given by
\begin{align}
    \frac{W}{dV dt} = \vec{\mathcal{E}} \cdot \vec{J}.
\end{align}
In turn, we are interested in the energy delivered per unit volume, which we calculate as
\begin{align} \label{eq:delivered_energy}
    \frac{W}{dV} = \int \vec{\mathcal{E}} \cdot \vec{J} \, dt.
\end{align}
In our simulations we use this delivered energy, measured in units of J$\cdot$m$^{-3}$, to estimate the breakdown probability inside silicon~\cite{Zhokhov2018}.\\

\begin{figure}
\includegraphics*[width=\linewidth]{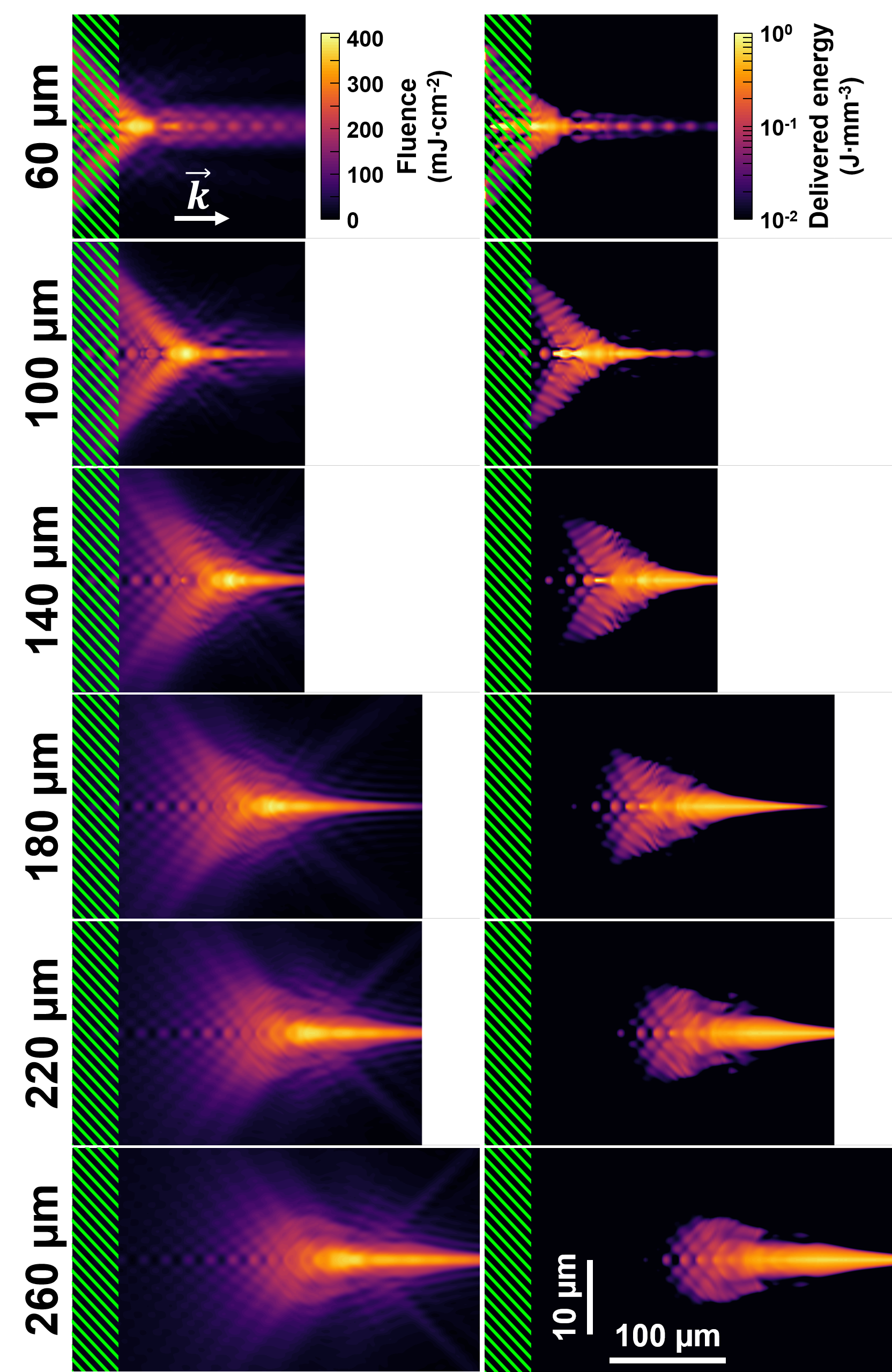}
\caption{\label{fig:FigAppendix} Simulated fluence and delivered energy distributions for different focusing depths indicated. The green shading marks the region below 40~\textmu m corresponding to surface damage in the experiments.}
\end{figure}

In order to provide an intuition, Fig.~\ref{fig:FigAppendix} shows the fluence and delivered energy distributions, calculated by Eq.~\eqref{eq:delivered_energy}, for different focusing depths. One can see that, compared to the fluence, the delivered energy distribution is much more localized in space (note the logarithmic scale for the delivered energy distributions in Fig.~\ref{fig:FigAppendix}). Following the experiments, we are interested only in bulk breakdown in silicon and, therefore, exclude from our analysis the energy delivered at depths below 40~\textmu m (marked by the green shading in Fig.~\ref{fig:FigAppendix}).

\section{\label{sec:AppendixB}Phenomenological model for transverse inscription}

Let us examine theoretically the conditions allowing transverse inscription of continuous lines inside silicon. To do so, we develop a phenomenological model relying on the laws of modification growth established in Sec.~\ref{sec:MultiPulseDamageGrowth} and geometrical considerations. Intuitively, high writing speeds should lead to a non-continuous (i.e, a discrete) line morphology consisting of individual single-pulse modifications next to each other. The discrete morphology thus solely exists when there is no interplay between individual modifications. In other words, the modification produced by one pulse does not provoke any shielding nor scattering effect during the subsequent irradiation.\\

\begin{figure}[b]
\includegraphics*[width=\linewidth]{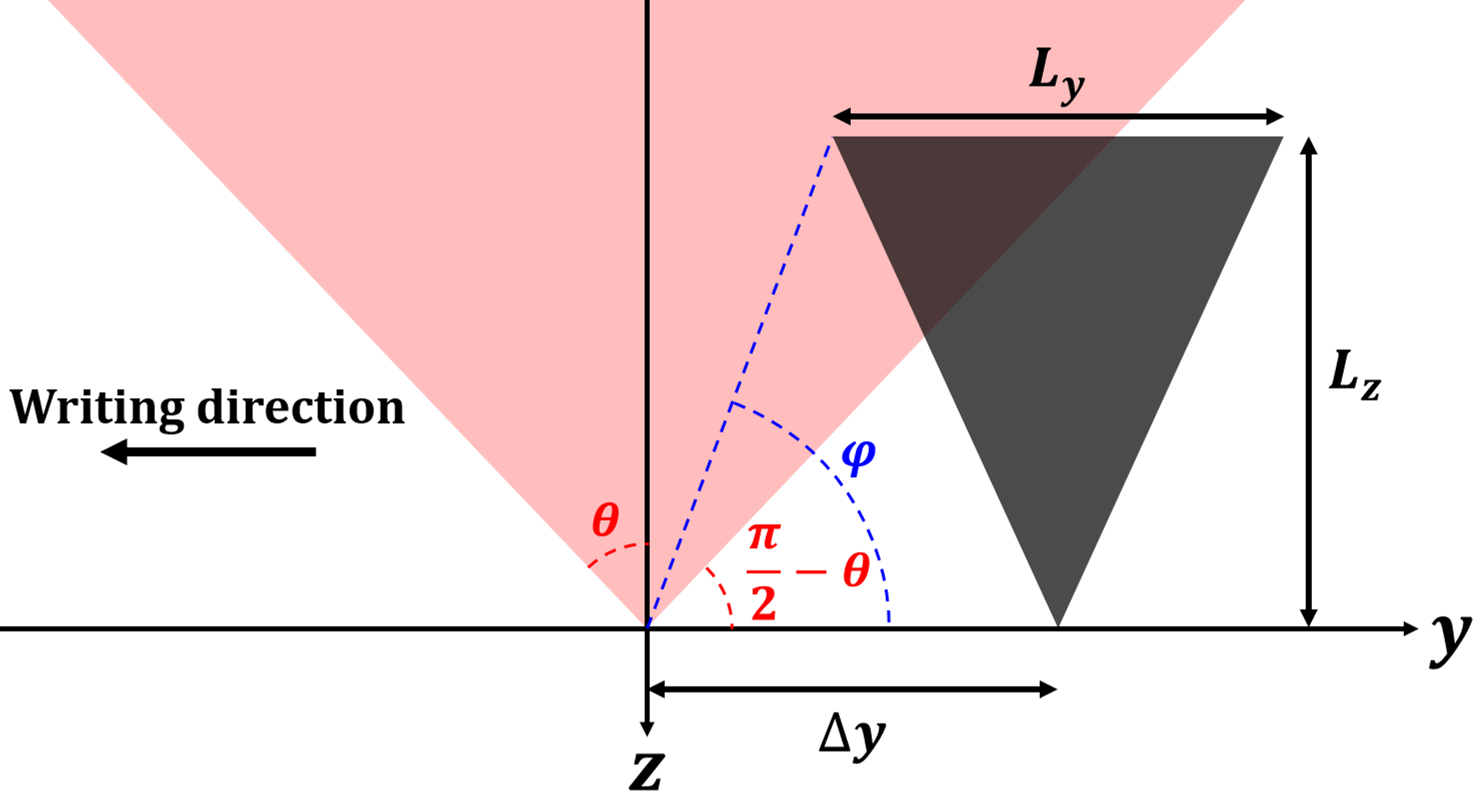}
\caption{\label{fig:FigWritingModel} Geometrical considerations associated with the phenomenological model for transverse inscription. The modification (in black) is produced by a pulse preceding the current irradiation (in red).}
\end{figure}

Without limiting the generality of the model, let us consider a sample movement---and, thus, an inscription---along the $y$-axis, as schematically depicted in Fig.~\ref{fig:FigWritingModel} where the laser irradiation (in red) is subsequent to a laser-produced modification (in black). The two consecutive irradiations are spatially separated by a distance $\Delta y=v/\Omega_{0}$, where $v$ is the writing speed, and $\Omega_{0}=100$ Hz is the laser repetition rate. In this figure, we introduce the angle $\varphi$ between the focal plane and the point of the modification with the closest distance to the optical axis $z$. The condition for discrete line inscription is thus
\begin{equation}
\label{eq:model1}
\varphi \le \frac{\pi}{2}-\theta,
\end{equation}
where $\theta=\text{arcsin}(\text{NA}/n_{0})$ is the maximum half-angle of the cone of light, $\text{NA}$ is the numerical aperture of the focusing lens, and $n_{0}$ is the refractive index of silicon at the laser wavelength. The angle $\varphi$ reads
\begin{equation}
\label{eq:model2}
\varphi = \text{arctan} \left( \frac{L_{z}}{\Delta y - L_{y}/2} \right).
\end{equation}
Here, $L_{z}$ and $L_{y}$ are the maximum lengths of the modification along $z$ and $y$, respectively. In the framework of our study, as established in Sec.~\ref{sec:MultiPulseDamageGrowth} for modification growth at the same position, these lengths both scale logarithmically with the number of pulses as
\begin{equation}
\label{eq:model3}
L_{y,z}=A_{y,z} \text{ln}(N)+B_{y,z},
\end{equation}
where $A_{y,z}$ and $B_{y,z}$ are the growth coefficient and the length of a single-pulse modification along $y$ and $z$, respectively, and $N \ge 1$ is the number of pulses applied during the growth sequence. In the case of transverse laser inscription where the focus position is continuously moved perpendicular to the laser, the number of pulses per point can be defined as $N=2 w_{0} \Omega_{0}/v$, where $w_{0}$ is the beam waist. Recalling that $\text{tan}(\pi/2-x)=1/\text{tan}(x)$, the combination of Eqs. (\ref{eq:model1}) and (\ref{eq:model2}) yields
\begin{equation}
\label{eq:model4}
\frac{L_{z}}{v/\Omega_{0}-L_{y}/2} \le \frac{1}{\text{tan} \left( \text{arcsin}(\text{NA}/n_{0}) \right)}.
\end{equation}

Finally, a condition on the writing speed $v$ can be expressed by combining Eqs. (\ref{eq:model3}) and (\ref{eq:model4}). This condition takes the form
\begin{equation}
\label{eq:model5}
\alpha v + \beta\text{ln}(v) \ge \gamma
\end{equation}
where
\begin{equation}
\label{eq:model6}
\begin{split}
\alpha = & 2/\Omega_{0} , \\
\beta = & A_{y}+2A_{z} \text{tan} \left( \text{arcsin}(\text{NA}/n_{0}) \right) , \\
\gamma = & A_{y} \text{ln}(2w_{0} \Omega_{0})+B_{y} \\
& +2(A_{z} \text{ln}(2w_{0} \Omega_{0})+B_{z})\text{tan}(\text{arcsin}(\text{NA}/n_{0}) . \\
\end{split}
\end{equation}
The threshold minimum speed $v_{\text{min}}$ for discrete line inscription appears in Eq.~(\ref{eq:model5}) when the left- and the right-hand sides are equal. In this case, the analytical solution is:
\begin{equation}
\label{eq:model7}
v_{\text{min}}=\frac{\beta}{\alpha} W\left( \frac{\alpha}{\beta} \text{exp} \left( {\frac{\gamma}{\beta}} \right) \right).
\end{equation}
Here, the notation $W$ denotes the Lambert $W$-function \cite{Corless1996}. Inserting in Eq.~(\ref{eq:model6}) the set of parameters determined experimentally in Sec.~\ref{sec:MultiPulseDamageGrowth}, we find a threshold minimum speed value for discrete line inscription $v_{\text{min}} \approx 262$ \textmu m$\cdot$s$^{-1}$.

\bibliography{references}

\end{document}